\documentclass[12pt]{article}
\usepackage{a4wide}
\usepackage{epsfig}
\usepackage{color}

\newcommand{\slk}{/\kern-6pt k}
\newcommand{\sll}{/\kern-4pt l}
\newcommand{\slp}{p\kern-5pt/}
\newcommand{\slq}{q\kern-5.5pt/}
\newcommand{\sls}{s\kern-5.5pt/}

\newcommand{\nn}{\nonumber\\}
\newcommand{\oone}{\hbox{$1\kern-2.5pt\hbox{\rm l}$}}
\newcommand{\ssigma}{\hbox{$\kern2.5pt\vrule height4pt\kern-2.5pt\sigma$}}

\newcommand{\GeV}{{\rm\,GeV}}

\newcommand{\Li}{\mathop{\rm Li}\nolimits}
\newcommand{\imag}{\mathop{\rm Im}\nolimits}

\newcommand{\slell}{/\kern-5pt\ell}

\begin{document}

\thispagestyle{empty} 
\begin{flushright}
MITP/17-084\\
\end{flushright}
\vspace{0.5cm}

\begin{center}

{\Large\bf Positivity bound on the imaginary part\\
  of the right-chiral tensor coupling \boldmath{$g_R$}\\[12pt]
  in polarized top quark decay}\\[1.3cm]
{\large S.~Groote$^1$ and J.G.~K\"orner$^2$ }\\[1cm]
$^1$ F\"u\"usika Instituut, Loodus- ja Tehnoloogiavaldkond,\\[.2cm]
  Tartu \"Ulikool, Wilhelm Ostwaldi 1, EE-50411 Tartu, Estonia\\[7pt]
  $^2$ PRISMA Cluster of Excellence, Institut f\"ur Physik,  \\[.2cm]
  Johannes-Gutenberg-Universit\"at, D-55099 Mainz, Germany\\[7pt]
 \end{center}

\vspace{1cm}
\begin{abstract}\noindent
We derive a positivity bound on the right-chiral tensor coupling $\imag g_R$
in polarized top quark decay by analyzing the angular decay distribution of
the three-body polarized top quark decay $t(\uparrow)\to b+\ell^+ +\nu_\ell$
in next-to-leading order QCD. We obtain the bound
$-0.0420 \le \imag g_R \le 0.0420$.
\end{abstract}

\newpage\noindent
The general matrix element for the decay $t \to b+W^+$ including the leading
order (LO) standard model (SM) contribution is usually written as (see e.g.\
Ref.~\cite{Bernreuther:2008us}) 
\begin{equation}
\label{genme}
M_{tbW^+}=-\frac{g_W}{\sqrt{2}}\varepsilon^{\mu\,\ast} \bar u_b
\Big[(V^\ast_{tb}+f_L)\gamma_\mu P_L +f_R \gamma_\mu P_R
  +\frac{i\sigma_{\mu\nu}\,q^\nu}{m_W}\Big(g_L\,P_L+g_R\,P_R\Big)\Big],
\end{equation}
where $P_{L,R}=(1\mp\gamma_5)/2$. The SM structure of the $tbW^+$ vertex is
obtained by dropping all terms except for the contribution proportional to
$V^\ast_{tb}\sim1$.

\begin{figure}\begin{center}
\epsfig{figure=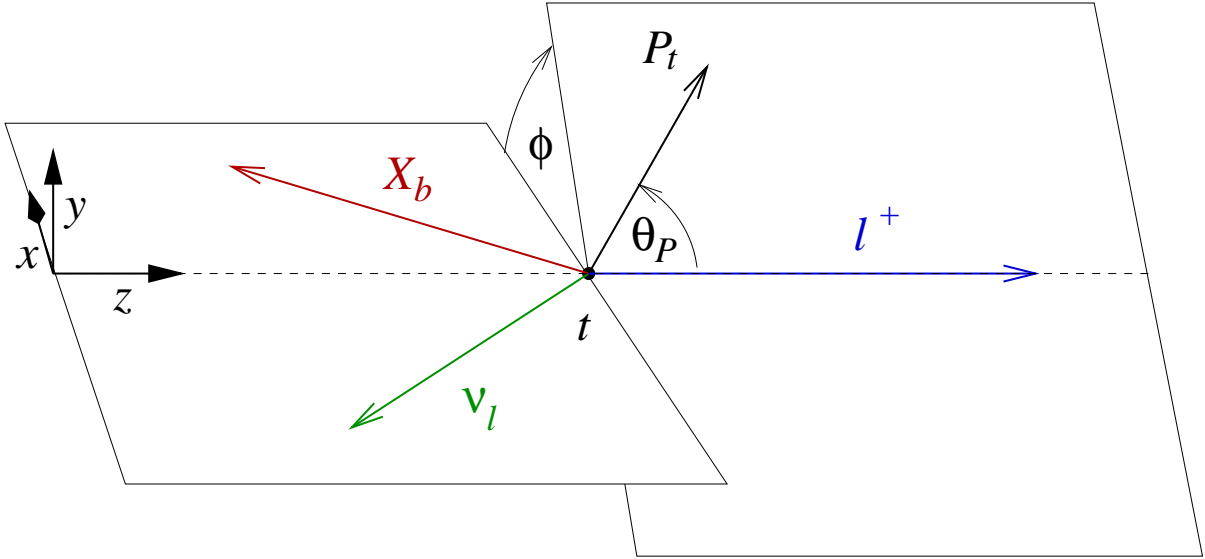, scale=0.8}
\caption{\label{D-azi}Definitions of polar and azimuthal angles for the
process $t \to b+W^+(\to \ell^++\nu_\ell)$}
\end{center}\end{figure}

The angular decay distribution for polarized top quark decay
$t(\uparrow)\to b+\ell^+ +\nu_\ell$ in the top quark rest frame is given by
\begin{eqnarray}
\label{angdis}
\frac{d\Gamma}{d\cos\theta d\phi}&=&
A+BP_t\cos\theta_P+CP_t\sin\theta_P\cos\phi+DP_t\sin\theta_P\sin\phi
\end{eqnarray}
which corresponds to the decay distribution introduced in
Refs.~\cite{Korner:1998nc,Groote:2006kq} augmented by the last $T$-odd term.
At LO of the SM one has $A=B$ and $C=D=0$. The second azimuthal term
proportional to $D$ corresponds to a $T$-odd contribution. This can be seen by
rewriting the angular factor as a triple product according to
\begin{equation}
\sin\theta_P\sin\phi=\hat p_\ell \cdot (\hat p_b \times \hat s_t)
\end{equation}
where (see Fig.~\ref{D-azi})
\begin{equation}
\hat p_\ell=(0,0,1) \qquad \hat p_b=(\sin\theta_b,0,\cos\theta_b)
\qquad \hat s_t=(\sin\theta_P \cos\phi,\cos\theta_P \sin\phi,\cos\theta_P)
\end{equation}
Let us repeat the arguments presented in Ref.~\cite{Groote:2006kq} that led us
to the conclusion that the equality $A=B$ already implies the vanishing of the
$T$-even azimuthal contribution $C$ confirming the LO result $C=0$. Consider
Eq.~(\ref{angdis}) for $P_t=1$ and $\sin\phi=0$, i.e.\ $\cos\phi=\pm 1$. Then
factor out the unpolarized rate term $A$. Assume first that $C/A$ is positive
and choose $\cos\phi=-1$. Expand the trigonometric functions around
$\theta_P=\pi$ for positive values of $\delta$, i.e.\
$\cos(\pi-\delta)\approx-1+\frac12\delta^2$ and
$\sin(\pi-\delta)\approx \delta$. The differential rate is then proportional to
\begin{equation}
1+\cos\theta_P-\frac{C}{A}\sin\theta_P\approx\frac12\delta^2
-\frac{C}{A}\delta=\frac{1}{2}\delta\left(\delta-\frac{2C}{A}\right).
\end{equation}
The differential rate can be seen to be negative for $\delta$ in the interval
$[0,2C/A]$. The interval can be shrunk to zero by setting $2C/A=0$, i.e.\ by
setting $C=0$. If $C/A$ is assumed to be negative, one chooses $\cos\phi=+1$
and again arrives at the conclusion $2C/A=0$ by expanding the trigonometric
functions around $\theta_P=\pi$.

The same chain of arguments but this time with $\cos\phi=0$ leads to the LO
positivity constraint for the $T$-odd structure, $D=0$.

At next-to-leading order (NLO) of QCD one no longer has $A=B$. However, the
relative difference $(A-B)/A$ is quite small which, as we will see, in turn
implies useful positivity constraints for the $T$-odd structure coefficient
$D$. As concerns the $T$-even azimuthal structure, the NLO corrections to the
LO result $C=0$ are so small that the positivity of the differential rate is
not endangered~\cite{Groote:2006kq}.

We now derive the NLO positivity constraint for the $T$-odd structure
coefficient $D$. We shall work in the approximation $m_b=0$ which implies that
the coupling terms $f_R$ and $g_L$ in Eq.~(\ref{genme}) are zero. The NLO
forms of the integrated $m_b=0$ rates are listed in
Refs.~\cite{Czarnecki:1990pe,Czarnecki:1993gt,Czarnecki:1994pu}. They read
\begin{eqnarray}
\frac{d\Gamma}{d\cos\theta d\phi}&=& (A^{(0)}+A^{(1)})\Big( 1+
\frac{A^{(0)}+B^{(1)} }{A^{(0)}+A^{(1)}}P_t\cos\theta_P \nn
&&+\frac{C}{A^{(0)}+A^{(1)}}P_t\sin\theta_P \cos\phi
+\frac{D}{A^{(0)}+A^{(1)}}P_t\sin\theta_P \sin\phi \Big),
\end{eqnarray}
where
\begin{eqnarray}
\frac{A^{(1)}}{A^{(0)}}&=&\frac{\alpha_sC_F}{4\pi}\,\frac1{(1-x^2)^2(1+2x^2)}
\bigg( (1-x^2)(5+9x^2-6x^4) \nn
&&-2(1-x^2)^2(1+2x^2)\Big[\frac{2\pi^2}{3} +4\ln(1-x^2)\ln x
  +4\Li_2(x^2) \Big]\nn
&&-8x^2(1+x^2)(1-2x^2)\ln x -2(1-x^2)^2(5+4x^2)\ln(1-x^2)\bigg)\nn
&=&-0.0846955,
\end{eqnarray}
and
\begin{eqnarray}
\frac{B^{(1)}}{A^{(0)}}&=&\frac{\alpha_sC_F}{4\pi}\,\frac1{(1-x^2)^2(1+2x^2)}
\bigg( -(1-x^2)(15-x^2+2x^4) \nn
&&+(1-x^2)(1+x^2+4x^4)\frac{2\pi^2}{3} -2(1-x^2)^2(5+4x^2)\ln(1-x^2)\nn
&&-16x^2(2+x^2-x^4)\ln x 
-16(1-x^2)(2+2x^2-x^4)\ln x \ln (1-x^2)\nn
&&-4(1-x^2)(5+5x^2-4x^4)\Li_2(x^2)\bigg) \nn
&=&-0.0863048,
\end{eqnarray}
where $x=m_W/m_t$. Here we have also listed the numerical values for the two
ratios using $\alpha_s(m_t)=0.1062$, $m_t=173.21\GeV$ and
$m_W=80.385\GeV$~\cite{Patrignani:2016xqp}. The ratio expressions
$A^{(1)}/A^{(0)}$ and $B^{(1)}/A^{(0)}$ have been rechecked in
Ref.~\cite{Groote:2006kq}. Reference~\cite{Groote:2006kq} also contains results
on the azimuthal rate coefficient $C$. This coefficient, however, will be of
no concern in the derivation of the positivity bounds for the $T$-odd rate
coefficient $D$. In fact, setting $sin\phi=\pm1$ will eliminate the
contribution of $C$. This will be our choice.

Next we must determine the contribution of the imaginary part of the coupling
factor $g_R$ to the $T$-odd azimuthal rate term $D$. The relevant contribution
arises from the interference of the coupling factor $g_R$ with the Born term
contribution. It is for this reason that there is no $\imag f_L$ contribution
to the $T$-odd rate coefficient $D$ since the coupling term is
self-interfering. After some algebra one finds
\begin{equation}
\frac{D}{A^{(0)}}=\frac{3\pi(1-x^2)}{4(1+2x^2)}\imag g_R
\end{equation}
where we have only kept the contribution linear in $\imag g_R$. Further, we
assume $\imag g_R $ to be positive, and set $P_t=1$ and $\sin\phi=-1$. We
expand around $\theta_P=\pi$ for small positive values of $\delta$ which gives
$\cos(\pi-\delta)=-1+\frac{1}{2}\delta^2$ and $\sin(\pi-\delta)=\delta$ to
obtain
\begin{equation}\label{WthetaP}
W(\theta_P)\sim 1+(1-\Delta)\cos\theta_P
-\frac{D}{A^{(0)}+A^{(1)}}\sin\theta_P
=\Delta-\frac{D\delta}{A^{(0)}+A^{(1)}}+\frac{1-\Delta}2\delta^2,
\end{equation}
where we have defined the small quantity
\begin{equation}
\Delta=\frac{A^{\rm NLO}-B^{\rm NLO}}{A^{\rm NLO}}
  =\frac{A^{(1)}-B^{(1)}}{A^{(0)}+A^{(1)}}
=\frac{A^{(1)}-B^{(1)}}{A^{(0)}(1+A^{(1)}/A^{(0)})},
\end{equation}
keeping in mind that $A^{(0)}=B^{(0)}$. Numerically one has $\Delta=0.001758$
where the small difference to the numerical results in
Ref.~\cite{Groote:2006kq} results from having used updated values
$m_W=80.385\GeV$ and $m_t=173.21\GeV$~\cite{Patrignani:2016xqp}.

The rate proportional to $W(\theta_P)$ in Eq.~(\ref{WthetaP}) becomes negative
if the contribution proportional to $\imag g_R$ becomes larger than the
remaining terms. However, this is no longer the case if the quadratic
equation~(\ref{WthetaP}) in $\delta$ has no real-valued zeros. The pertinent
condition for the discriminant reads
\begin{equation}
\frac{3\pi(1-x^2)}{4(1+2x^2)}|\imag g_R|
  \le\sqrt{2\Delta(1-\Delta)}\left(1+\frac{A^{(1)}}{A^{(0)}}\right).
\end{equation}
Numerically one obtains
\begin{equation}                                                    
\imag g_R \le\,0.0420
\end{equation}
The same chain of arguments, but now for negative values of $\imag g_R$ and
$\sin\phi=1$ leads to $\imag g_R \ge\,-0.0420$ such that one has the two-sided
constraint
\begin{equation}
\label{bound0}                                                    
\imag g_R \in [-0.0420,\,0.0420]\,.
\end{equation}
The angle $\delta_0$ for which the quadratic form~(\ref{WthetaP}) becomes
degenerate can be calculated to be
$\delta_0=\sqrt{2\Delta/(1-\Delta)}=\pm 0.0593$. The corrections to the
expansion of $\cos(\pi - \delta)$ and $\sin(\pi - \delta)$ are of the
order $O(\delta^2=0.00352)$ and, therefore, quite small.

In Ref.~\cite{Fischer17} we have calculated the SM absorptive electroweak
contributions to $\imag g_R$ with the result $\imag g_R=-2.175\times 10^{-3}$
(see also Refs.~\cite{GonzalezSprinberg:2011kx,Arhrib:2016vts}). This value is
easily accommodated in the positivity bound~(\ref{bound0}).

The ATLAS Collaboration has recently published the bound~\cite{Aaboud:2017aqp}
\begin{equation}
\imag g_R \in [-0.18,\,0.06]
\end{equation}
based on the analysis of sequential polarized two-body top quark decays
$t(\uparrow) \to b + W^+ (\to \ell^+ + \nu)$.
A somewhat tighter bound has been published in Ref.~\cite{Aaboud:2017yqf}
using also sequential polarized two-body top quark decays. The bound reads  
\begin{equation}
\label{bound2}
\imag\left(\frac{g_R}{f_L}\right) \in [-0.07,\,0.06]
\end{equation}
which we translate into a bound on $\imag g_R$ by substituting the LO result
$f_L=1$ in Eq.~(\ref{bound2}). Both bounds are not far away from the
positivity bound on $\imag g_R$ derived in this paper. Using the same chain of
arguments one can establish the corresponding bound for $m_b\ne 0$. Using NLO
$m_b\ne 0$ results from Ref.~\cite{Czarnecki:1994pu} on the unpolarized and
polarized rate functions $A^{(1)}$ and $B^{(1)}$ we find that for $m_b=4.8\GeV$
the bound is marginally strengthened to $\imag g_R\in[-0.0418,0.0418]$. The
condition for obtaining this bound reads
\begin{equation}
\frac{3\pi(1-x^2+y^2)\sqrt\lambda}{4\left(\lambda+3x^2(1-x^2+y^2)\right)}
  |\imag g_R|\le\sqrt{2\Delta(1-\Delta)}\left(1+\frac{A^{(1)}}{A^{(0)}}\right),
\end{equation}
where $y=m_b/m_t$ and $\lambda=\lambda(1,x^2,y^2)=1+x^4+y^4-2x^2-2y^2-2x^2y^2$
is the K\"all\'en function. $A^{(0)}$, $A^{(1)}$ and the small quantity
$\Delta$ are evaluated for $m_b\ne 0$.

\subsection*{Acknowledgments}
We would like to thank J.~Mueller for discussions.
This work was supported by the Estonian Science Foundation under Grant
No.~IUT2-27. S.G.\ acknowledges the hospitality of the theory group THEP at
the Institute of Physics at the University of Mainz and the support of the
Cluster of Excellence PRISMA at the University of Mainz. 

\end{document}